\documentclass[11pt,twoside]{article}
\usepackage{asp2010}

\resetcounters

\begin{document}

\title{PTF/M-dwarfs: First Results From a Large New M-dwarf Planetary Transit Survey}
\author{Nicholas~M.~Law$^1$, Adam~L.~Kraus$^2$, Rachel~R.~Street$^3$, Tim~Lister$^3$, Avi~Shporer$^{3,4},$, Lynne~A.~Hillenbrand$^5$, and The Palomar Transient Factory Collaboration$^6$
\affil{$^1$Dunlap Fellow; Dunlap Institute for Astronomy and Astrophysics, University of Toronto, 50 St.\ George Street, Toronto M5S 3H4, Ontario, Canada}
\affil{$^2$Hubble Fellow; University of Hawaii-IfA, 2680 Woodlawn Drive, Honolulu, HI 96822, USA}
\affil{$^3$Las Cumbres Observatory Global Telescope Network, 6740 Cortona Dr. Suite 102, Goleta, CA 93117, USA}
\affil{$^4$Department of Physics, Broida Hall, University of California, Santa Barbara, CA 93106, USA}
\affil{$^5$California Institute of Technology, Department of Astrophysics, MC 249-17, Pasadena, CA 91125, USA}
\affil{$^6$http://www.astro.caltech.edu/ptf/}
}

\begin{abstract}
PTF/M-dwarfs is a 100,000-target M-dwarf planetary transit survey, a Key Project of the Palomar Transient Factory (PTF) collaboration. The survey is sensitive to Jupiter-radius planets around all of the target stars, and has sufficient precision to reach Neptunes and super-Earths for the best targets. The Palomar Transient Factory is a fully-automated, wide-field survey aimed at a systematic exploration of the optical transient sky. The survey is performed using a new 7.26 square degree camera installed on the 48 inch Samuel Oschin telescope at Palomar Observatory. Each 92-megapixel R-band exposure contains about 3,000 M-dwarfs usable for planet detection. In each PTF observational season PTF/M-dwarfs searches for Jupiter-radius planets around almost 30,000 M-dwarfs, Neptune-radius planets around approximately 500 M-dwarfs, and super-Earths around 100 targets. The full survey is expected to cover more than 100,000 targets over the next several years. Photometric and spectroscopic followup operations are performed on the Palomar 60-inch, LCOGT, Palomar 200-inch, MDM and Keck telescopes. The survey has been running since mid-2009. We detail the survey design, the survey's data analysis pipeline and the performance of the first year of operations.
\end{abstract}

\section{Introduction}

Thus far only a handful of planetary systems have been detected around stars with masses $< 0.5\rm{M_{\odot}}$, and very few lower-mass M-dwarfs have been searched for planets. The higher-mass M-dwarfs that have been probed up to now have a low Jupiter-mass planet companion frequency ($\sim$2\%; eg. \citet{Johnson2007}). However, theorists predict a large population of lower-mass planets around all masses of M-dwarfs \citep{Ida2005, Kennedy2008}, and recent microlensing and transit detection results (e.g. \citet{Sumi2010, Charbonneau2009}) suggest that superEarths and mini-Neptunes may be very common around sub-solar-mass stars.

PTF/M-dwarfs is a new transit detection survey targeted at 100,000 cool stars. The survey is a key project of the Palomar Transient Factory (PTF) collaboration (\citet{Law2009} \& \citet{Rau2009}). PTF is a new wide field, multiple cadence transient search hosted at the 1.2m Samuel Oschin Telescope at Palomar Observatory (hereafter referred to as P48). Commissioning of the new 7.26-square-degree camera system was completed in August 2009, and PTF science operations are ongoing \citep{Law2010}.

\begin{figure}[tb]
\plotone{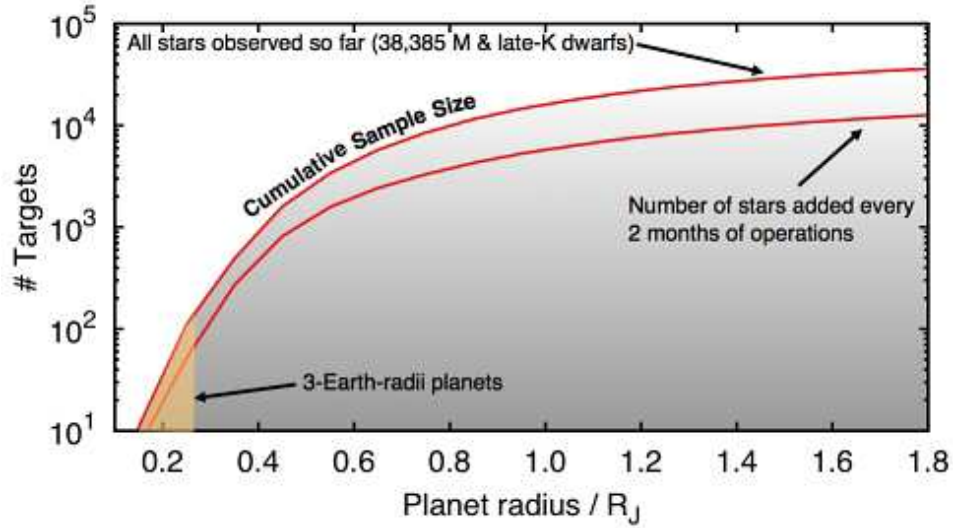}
\caption{\label{ps}The cumulative PTF/M-dwarfs sample size in the first 9 months of operations, as a function of detectable planet radius. The planet sensitivity of each target is calculated on the basis of the achieved RMS photometric precision for that target and its photometrically-estimated spectral type. In this figure we assume that a planet detection requires a 2-sigma detection in each of at least three datapoints during a transit. This assumption is a conservative estimate that is improved with the utilization of full phase-folding detection techniques. Observation window effects are not included here.}
\end{figure}

\begin{figure}[tb]
\plotone{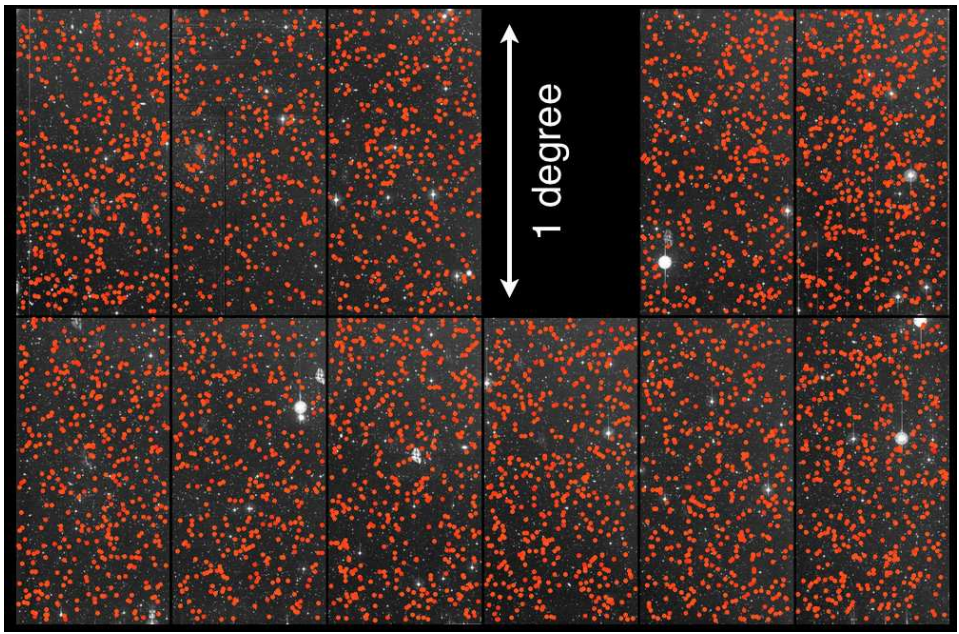}
\caption{\label{img}A typical PTF/M-dwarfs target field, with the planet-detection targets denoted by red points.}
\end{figure}

\section{Survey Setup, Observations and Data Reduction}

PTF/M-dwarfs is planned to complete transit-search observations of 100,000 M-dwarfs by 2012 ($\sim$40,000 targets have been observed up to Q4 2010; figure \ref{ps}). The survey is designed to be sensitive to giant planets in $<$ 15-day period orbits around M-dwarfs, although it is also capable of detecting Neptune-radius and even rocky planets around a subset of the targets. The PTF/M-dwarfs survey parameters are summarized in table \ref{tab}.

Each PTF 60-second exposure covers 7.26 square degrees with a limiting magnitude of $\rm{m_R}$$\approx$21. Our planet-detection sample includes all M-dwarf targets in our fields and brighter than R=19, where the achieved photometric stability is approximately 10\%. Each field contains approximately 3000 M-dwarfs suitable for planet detection (figure \ref{img}). The PTF/M-dwarfs survey covers four fields every two months at an 18-minute cadence, for a total of $\sim$12,000 new planet-search targets in 29.04 square degrees of sky. M-dwarf targets are selected based on fits to SDSS, USNO-B and 2MASS photometry. We reject giants from our sample by including only stars with significant proper motions noted in the SDSS-DR7 database.

We have used a detailed model of the M-dwarf galactic stellar distribution to determine our target densities and optimal pointing direction. Although the number of target stars could be increased by observing in the galactic plane, we have instead elected to target a galactic latitude of approximately 30 degrees, looking away from the galactic center, to 1) greatly reduce the probability of giant star interlopers; 2) improve our photometric precision by avoiding crowding; and 3) reduce the probability of blends: combinations of eclipsing systems and another nearby star that have low apparent eclipse depth and can look like planet transits. Wherever possible we target SDSS fields to take advantage of the accurate multi-color photometry and proper motions.

\begin{table}
\label{tab:camera_specs}

\begin{tabular}{ll}
\multicolumn{2}{l}{\bf P48 PTF camera specifications }\\
\hline
Telescope     & Palomar 48-inch (1.2m) Samuel Oschin \\
Camera field dimensions & 3.50 $\times$ 2.31 degrees\\
Camera field of view        & 8.07 square degrees \\
Light sensitive area        & 7.26 square degrees \\
Image quality        & 2.0 arcsec FWHM in median seeing \\
Filters              & $\rm{g^\prime}$ \& Mould-R; other bands available\\
CCD specs            & 2K$\times$4K MIT/LL 3-edge butted CCDs\\
Plate scale             &  1.01 arcsec / pixel\\
Readout noise        & $<$ 12 $\rm{e^-}$\\
Readout speed        & 35 seconds, entire 100 MPix array\\
Efficiency           & 66\% open-shutter (slew during readout)\\
Saturation Level     & $\rm{m_R}=14-15$ (seeing dependant)\\
Sensitivity (median)          & $\rm{m_R}$$\approx$21 in 60 s, 5$\sigma$\\
                               & $\rm{m_{g^\prime}}$$\approx$21.3 in 60 s, 5$\sigma$ \\
\vspace{0.25cm}\\

\multicolumn{2}{l}{\bf PTF/M-dwarfs survey characteristics}\\
\hline

Targets & Late-K, M and L dwarfs with $\rm{m_R} < 19$\\
Survey sky area & 29.04 square degrees every 2 months\\
Target locations & 20$^{\circ}$ $<$ galactic latitude $<$ $35^{\circ}$ \\
Targets covered & $\sim$12,000 every 2 months\\
Observations per night & 5 hours\\
Exposure time   & 60 seconds \\
Cadence & 18 minutes\\
Observation length & 1--3 months \\
Photometric precision per datapoint & 3 millimag (brightest targets) \\
                      & 10\% (faintest targets)\\
Followup & Photometric: Palomar 60-inch, LCOGT FTN \& FTS\\
         & Low-res spectroscopic: Palomar 200-inch, Lick Shane-3m\\
         & Radial velocity: Keck I / HIRES \\
\hline \vspace{0.25cm} \\
\end{tabular}

\caption{\label{tab}The specifications of the PTF Camera and the PTF/M-dwarfs survey.}
\end{table}

\subsection{Data Reduction}
Each 92MPix image from P48 is flat-fielded, debiased, and astrometrically calibrated by the PTF LBNL supernova search pipeline (for more details, see e.g. \citealt{Law2009}). The calibrated images are then passed through the custom PTF/M-dwarfs pipeline for source extraction, source matching, light curve generation, precision photometric calibration, and transit and eclipse detection.

The PTF/M-dwarfs pipeline is Python-based, although C modules are used for computationally intensive components. All calculations are done on a per-CCD basis, so each sky region considered covers 4096$\times$2048 pixels, or $1.14^{\circ}\times0.57^{\circ}$. Initial source extraction is performed by SExtractor \citep{Bertin1996} using radius-optimized aperture photometry with a locally-optimized background. The extracted sources are filtered to remove those close to bad pixels, diffraction spikes from bright stars, and other effects. The resulting list of photometric data points (typically 3000-6000 sources per chip each with several hundred epochs) is then merged into single light curves with an astrometric matching radius of 2.0 arcseconds; this is much larger than the typical astrometric precision produced by the PTF camera \citep{Law2010}.

\begin{figure}[tb]
\plotone{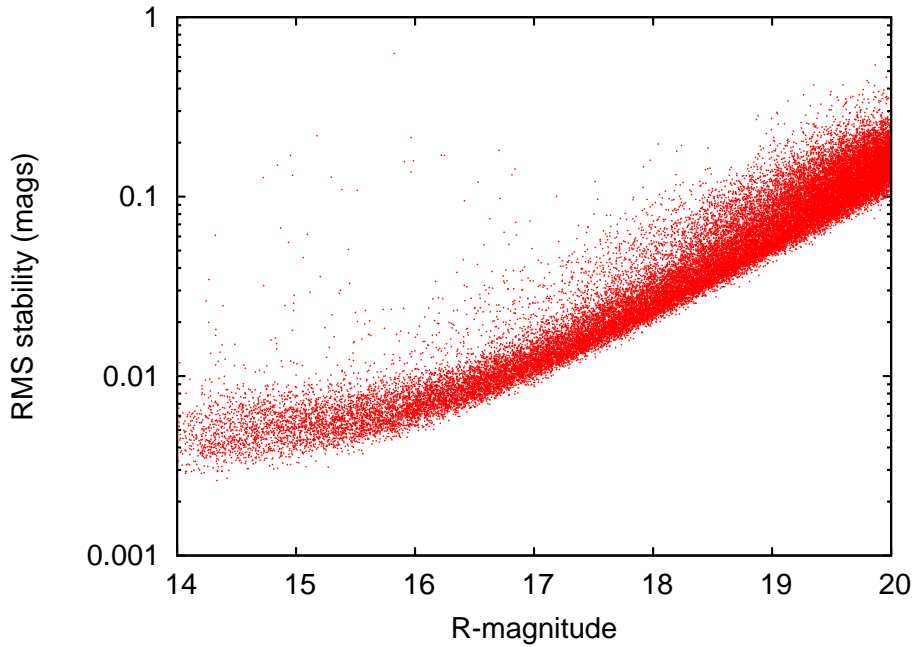}
\caption{\label{phot}The RMS photometric precision achieved by the PTF/M-dwarfs pipeline for a typical field over approximately one month. Each point is a single star in the field. Note the few stars with very high variability compared to others at similar magnitudes: these objects are astrophysically varying sources such as eclipsing binaries.}
\end{figure}

The photometric zeropoints for each epoch are initially estimated based on either SDSS or USNO-B1 photometry for bright stars in the field. This results in photometric precisions of a few percent. For the purposes of more precise photometric calibration, we remove sources detected as being highly variable in the initial light curves. The pipeline then optimizes the zeropoint of each epoch to minimize the median photometric variability of all the remaining sources, using a Nelder-Mead downhill simplex algorithm. This first optimization typically improves the long-term photometric stability to below the percent level.

The pipeline then filters the generated light curves, searching for epochs which produce anomalous photometry for a large fraction of the sources; those epochs are usually those affected by clouds, moonlight or some other effect which varies across the images. Typically 0-2\% of epochs are flagged by this process, and are removed from further consideration. A second iteration of variable-source removal and zeropoint optimization is then performed. The final zeropoints are applied to each original light curve. 

The pipeline's output photometric stability is typically a few millimags over periods of months for the brightest stars in our fields (figure \ref{phot}). The photometric precision is photon-limited for all sources fainter than $\rm{m_R}$$\sim$16, except in regions of crowding or nebulosity. Running on a 2.5 GHz quad-core desktop computer the pipeline typically processes a 300-epoch set of 11 chips (54 GB of image data) in less than 24 hours.

\begin{figure}[tb]
\plotone{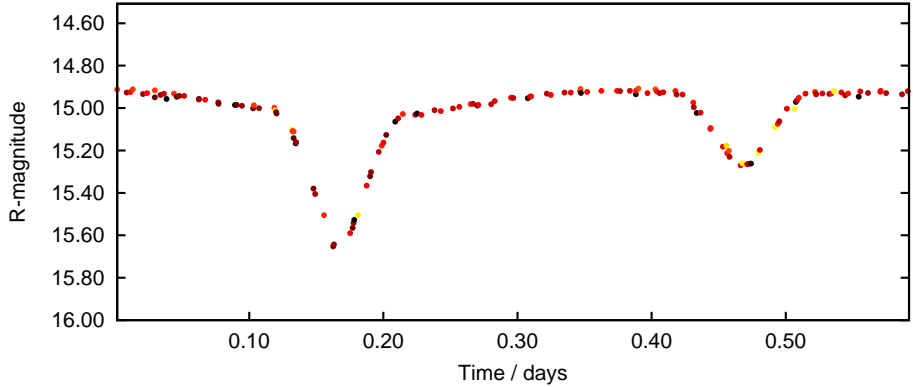}
\caption{\label{eclipse}An example P48 discovery light curve for a 15th-magnitude eclipsing system containing a K-dwarf and an M-dwarf. The light curve includes 30 nights of data folded to the system's 0.6025 day period. The points' colors correspond to their epoch, with the earliest data as black, data taken in the middle of the run as red, and the latest data as yellow. The median system magnitude is shown to guide the eye.}
\end{figure}

We search the calibrated light curves for eclipsing and transiting objects by: 1) surveying high-variability M-dwarf light curves by eye; 2) an automated search for obvious (multiple-sigma) dips in M-dwarf light curves; and 3) an automated box-least-squares \citep{Kovacks2002} search for periodic dips in all the light curves in our dataset. An example phase-folded light curve detected by our automated routines is shown in figure \ref{eclipse}.

\subsection{Candidate follow up}
After detection of a planetary candidate, the target is passed for follow up with 1) higher-cadence, high-precision photometry for eclipse confirmation and blend removal (our target stars are much redder than the median stars in the field so a blend is very likely to produce a color shift during eclipse as the relative contribution from the two stars changes); 2) low-resolution spectroscopy to accurately determine the target spectral type and to confirm our proper-motion-based dwarf/giant classification; and 3) final radial velocity confirmation with Keck I / HIRES.

Photometric follow up is performed on the Palomar 60-inch telescope and the 2m Faulkes North and Faulkes South telescopes. Low-resolution spectroscopy is undertaken on a variety of telescopes including the Palomar 200-inch and the Lick Observatory Shane 3m telescope. 

The survey's very red targets are bright enough at red wavelengths ($\rm{m_I} < 18.0$) to allow us to perform Keck HIRES followup at kilometre-per-second level precisions. This radial velocity precision is sufficient to either reject the possibility of a brown dwarf in the system at short periods, or directly detect close-in massive planets.

\section{Survey Status}

\begin{figure}[tb]
\plotone{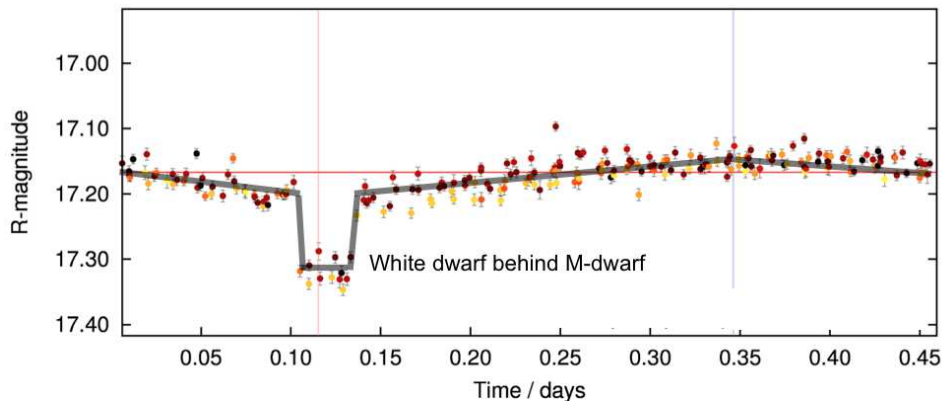}
\caption{\label{wd}An eclipsing white-dwarf / M-dwarf binary detected by PTF/M-dwarfs. Much of the photometric scatter is caused by stellar activity.  Keck/HIRES observations show the M-dwarf to have a radial velocity semiamplitude of $\approx$150 km/sec. We are currently engaged in a comprehensive study of this system to obtain precision masses and radii of the components.}
\end{figure}

In its first 9 months of operation the survey has obtained 200-300 epochs of data on 10 target fields. In total the data in hand covers 38,385 late-K and M-dwarfs with sensitivity to giant planets and good detection probabilities for periods less than about 10 days. As described in figure \ref{ps}, as well as giant planets the survey is sensitive to Neptune-radius planets around several thousand targets and rocky planets around approximately one hundred targets (increasing to several hundred targets once the survey is complete).

PTF/M-dwarfs has discovered several signals with eclipse amplitudes consistent with a Jupiter-radius companion. Radial velocity follow-up is underway for those planet candidates; none have yet been confirmed. We have also discovered 28 eclipsing binaries with two M-dwarf components, and we are obtaining precision masses and radii for their components. The survey has also detected more exotic systems such as white-dwarf / M-dwarf binaries (figure \ref{wd}).

Once complete in 2012, PTF/M-dwarfs will have obtained several-hundred-epoch light curves for around 100,000 M-dwarf targets, with sufficient photometric precision to detect transiting giant planets. As well as exoplanet detections, this dataset will also enable a wide range of other science -- for example stellar activity and rotation studies (e.g. Covey et al. 2010, these proceedings), proper motion surveys, and searches for exotic binary systems.

\acknowledgements This research has made use of NASA's Astrophysics Data System Bibliographic Services, the SIMBAD database, operated at CDS, Strasbourg, France, and the VizieR database of astronomical catalogs. 

\bibliography{law_n}

\end{document}